%% file: new.tex
\documentclass[twocolumn]{aastex63}
\usepackage{savesym}
\savesymbol{tablenum}
\usepackage{siunitx}
\restoresymbol{SIX}{tablenum}

\usepackage[utf8]{inputenc}
\usepackage{graphicx}
\usepackage{amsmath}
\usepackage{siunitx,booktabs}
\usepackage{array}
\usepackage{longtable}
\usepackage{lineno}
\usepackage[encapsulated]{CJK}

\newcommand{\oiiiboth}{[O \textsc{iii}]$\lambda\lambda4959,5008$}
\newcommand{\oiiifive}{[O \textsc{iii}]$\lambda5008$}

\usepackage{xspace}

\begin{document}
\include{commands}
\shorttitle{}

\shortauthors{Lewis et al.}

\title{The Mass--Metallicity Relation and its Observational Effects at $z\sim$3-6}

\author[0000-0003-0695-4414]{Zach Lewis}\thanks{NSF Graduate Research Fellow}
\affiliation{Department of Astronomy, University of Wisconsin-Madison, Madison, WI 53706, USA}

\author[0000-0003-0695-4414]{Michael V. Maseda}
\affiliation{Department of Astronomy, University of Wisconsin-Madison, Madison, WI 53706, USA}

\author[0000-0002-2380-9801]{Anna de Graaff}
\thanks{Clay Fellow}
\affiliation{Center for Astrophysics $|$ Harvard \& Smithsonian, 60 Garden St., Cambridge MA 02138 USA}
\affiliation{Max-Planck-Institut f\"ur Astronomie, K\"onigstuhl 17, D-69117, Heidelberg, Germany}

\author[0000-0001-6755-1315]{Joel Leja}
\affiliation{Department of Astronomy \& Astrophysics, The Pennsylvania State University, University Park, PA 16802, USA}
\affiliation{Institute for Computational \& Data Sciences, The Pennsylvania State University, University Park, PA 16802, USA}
\affiliation{Institute for Gravitation and the Cosmos, The Pennsylvania State University, University Park, PA 16802, USA}

\author[0000-0001-9269-5046]{Bingjie Wang (\begin{CJK*}{UTF8}{gbsn}王冰洁\ignorespacesafterend\end{CJK*})}
\thanks{NHFP Hubble Fellow}
\affiliation{Department of Astrophysical Sciences, Princeton University, Princeton, NJ 08544, USA}

\author[0000-0003-4996-9069]{Hans-Walter Rix}
\affiliation{Max-Planck-Institut f\"ur Astronomie, K\"onigstuhl 17, D-69117, Heidelberg, Germany}

\author[0000-0002-2446-8770]{Ian McConachie}
\affiliation{Department of Astronomy, University of Wisconsin-Madison, Madison, WI 53706, USA}

\author[0000-0001-7151-009X]{Nikko J. Cleri}
\affiliation{Department of Astronomy \& Astrophysics, The Pennsylvania State University, University Park, PA 16802, USA}
\affiliation{Institute for Computational \& Data Sciences, The Pennsylvania State University, University Park, PA 16802, USA}
\affiliation{Institute for Gravitation and the Cosmos, The Pennsylvania State University, University Park, PA 16802, USA}

\author[0000-0001-5063-8254]{Rachel Bezanson}
\affiliation{Department of Physics \& Astronomy and PITT PACC, University of Pittsburgh, Pittsburgh, PA 15260, USA}

\author[0000-0002-3952-8588]{Leindert A. Boogaard}
\affiliation{Leiden Observatory, Leiden University, P.O. Box 9513, 2300 RA Leiden, The Netherlands}

\author[0000-0003-2680-005X]{Gabriel Brammer}
\affiliation{Cosmic Dawn Center (DAWN), Denmark}
\affiliation{Niels Bohr Institute, University of Copenhagen, Jagtvej 128,
K{\o}benhavn N, DK-2200, Denmark}

\author[0000-0002-5612-3427]{Jenny E. Greene}
\affiliation{Department of Astrophysical Sciences, Princeton University, Princeton, NJ 08544, USA}

\author[0000-0002-3301-3321]{Michaela Hirschmann}
\affiliation{Institute of Physics, Lab for galaxy evolution, EPFL, Observatoire de Sauverny, Chemin Pegasi 51, 1290 Versoix, Switzerland}

\author[0000-0003-1561-3814]{Harley Katz}
\affiliation{Sub-department of Astrophysics, University of Oxford, Oxford OX1 3RH, UK}

\author[0000-0002-2057-5376]{Ivo Labb\'e}
\affiliation{Centre for Astrophysics and Supercomputing, Swinburne University of Technology, Melbourne, VIC 3122, Australia}

\author[0000-0003-2871-127X]{Jorryt Matthee}
\affiliation{Institute of Science and Technology Austria (ISTA), Am Campus 1, 3400 Klosterneuburg, Austria}

\author[0000-0001-8367-6265]{Tim B. Miller}
\affiliation{Center for Interdisciplinary Exploration and Research in Astrophysics (CIERA), Northwestern University, IL 60201, USA}

\author[0000-0003-3997-5705]{Rohan P. Naidu}
\thanks{NHFP Hubble Fellow}
\affiliation{MIT Kavli Institute for Astrophysics and Space Research, Cambridge, MA 02139, USA}

\author[0000-0001-5851-6649]{Pascal A. Oesch}
\affiliation{Department of Astronomy, University of Geneva, Chemin Pegasi 51, 1290 Versoix, Switzerland}
\affiliation{Cosmic Dawn Center (DAWN), Copenhagen, Denmark}

\author[0000-0003-4075-7393]{David J. Setton}
\thanks{Brinson Prize Fellow}
\affiliation{Department of Astrophysical Sciences, Princeton University, Princeton, NJ 08544, USA}

\author[0000-0002-1714-1905]{Katherine A. Suess}
\affiliation{Department for Astrophysical \& Planetary Science, University of Colorado, Boulder, CO 80309, USA}

\author[0000-0001-8928-4465]{Andrea Weibel}
\affiliation{Department of Astronomy, University of Geneva, Chemin Pegasi 51, 1290 Versoix, Switzerland}

\author[0000-0001-7160-3632]{Katherine E. Whitaker}
\affiliation{Department of Astronomy, University of Massachusetts, Amherst, MA 01003, USA}
\affiliation{Cosmic Dawn Center (DAWN), Copenhagen, Denmark}

\author[0000-0003-2919-7495]{Christina C.\ Williams}
\affiliation{NSF's National Optical-Infrared Astronomy Research Laboratory, 950 North Cherry Avenue, Tucson, AZ 85719, USA}
\affiliation{Steward Observatory, University of Arizona, 933 North Cherry Avenue, Tucson, AZ 85721, USA}

\begin{abstract}
    The correlation between galaxy stellar mass and gas-phase metallicity, known as the mass--metallicity relation (MZR), gives key insights into the processes that govern galaxy evolution. However, unquantified observational and selection biases can result in systematic errors in attempts to recover the intrinsic MZR, particularly at higher redshifts. We characterize the MZR at $z\sim3-6$ within a fully Bayesian framework using JWST NIRSpec spectra of 193 galaxies from the RUBIES survey. We forward model the observed mass--metallicity surface using \texttt{prospector}-generated spectra to account for two selection biases: the survey selection function and success in observing high signal-to-noise emission lines. We demonstrate that the RUBIES selection function, based on F444W magnitude and F150W-F444W color, has a negligible effect on our measured MZR. A correct treatment of the non-Gaussian metallicity uncertainties from strong-line calibrations lowers the derived MZR normalization by 0.2 dex and flattens the slope by $\sim$20\%; forward-modeling the effect of emission line observability steepens the slope by $\sim$15\%. Both of these biases must be taken into account in order to properly measure the intrinsic MZR. This novel forward modeling process motivates careful consideration of selection functions in future surveys, and paves the way for robust, high-redshift chemical enrichment studies that trace the evolution of the mass--metallicity relation across cosmic time.
\end{abstract}


\section{Introduction}

The relative chemical abundances of the interstellar medium (ISM), or gas-phase metallicity (hereafter metallicity), of a galaxy is sensitive to many key and unknown processes of galaxy evolution. The ISM is enriched by star formation, diluted by inflows of pristine gas or outflows of metal-rich gas driven by active galactic nuclei (AGN) or stellar winds, and is sensitive to mixing processes within a galaxy (see, e.g., \citealt{tumlinson2017} and \citealt{maiolino2019} and references therein). 

The metallicity of a galaxy is also tightly correlated with its stellar mass (\citealt{tremonti2004}, \citealt{curti2020}). Infalling gas is turned into stars, increasing the stellar mass of the galaxy. These stars enrich the interstellar medium of a galaxy upon their death, increasing the metallicity. In the local universe, the MZR follows a power law at low stellar masses, before turning over and appearing to saturate at higher stellar masses, approaching an asymptotic metallicity \citep[e.g.,][]{tremonti2004, andrews2013, curti2020}. This form is also in place at intermediate redshifts ($z\sim$0.5-1.5; \citealt{perezmontero2009}, \citealt{zahid2011}, \citealt{topping2021}, \citealt{lewis2024}) and beyond cosmic noon (\citealt{sanders2018}, \citealt{sanders2021}, \citealt{papovich2022}). This saturation may represent the chemical balance between the oxygen locked up in long-lived low-mass stars and that created by massive stars \citep{zahid2013}, or could be due to massive galaxies no longer forming stars in situ, but rather growing through accretion of older stars. 

It remains to be seen whether this form is in place at redshifts nearing the early Universe, or, if not, when and why this relation sets in. JWST has enabled the measurement of galaxy gas-phase metallicity past $z\sim10$ (\citealt{langeroodi2022}, \citealt{nakajima2023}, \citealt{sarkar2024}, \citealt{morishita2024}, \citealt{curti2024}). These studies, while pushing the boundaries of our understanding of early-universe chemical enrichment, are often the observed medians of samples derived from surveys with complicated selection functions, and do not address the bias introduced by the inability to measure weak emission line fluxes, for example. 

At these higher redshifts ($z\sim$4-6), it becomes especially important that observed mass--metallicity relations are robust. First, it is advantageous to perform metallicity studies with surveys that are representative in mass and metallicity parameter space. All surveys are subject to selection effects, but a selection function that is easily parameterized can be combined with a catalog of artificial galaxy realizations to provide a way to correct for the regimes in which an observed MZR may not be representative of the true galaxy population.

Second, the ability to produce strong lines that are tracers of metallicity is a function of metallicity itself, through star formation rate. For example, more highly star-forming galaxies at fixed stellar mass are likely to exhibit lower metallicities, as well as stronger emission lines \citep{mannucci2010}. Metallicities are more easily measured in galaxies with stronger emission lines. If unaccounted for, this may lead to a measurement of the MZR that is flatter than would be measured for the true galaxy population.

Understanding galaxy chemical evolution thus necessitates the measurement of robust observed mass--metallicity relations. Beyond being useful in their own right as a metric to understand the baryon cycle and its changes across cosmic time, cosmological simulations often use observed MZRs as a benchmark scaling relation to which their subgrid physical models are tuned (\citealt{torrey2019}, \citealt{dave2017}). Since simulated MZRs are not subject to selection effects or metallicity measurement biases, tuning to a biased observed MZR would result in incorrect subgrid prescriptions. Additionally, metallicities derived from emission lines differ from ``intrinsic'' metallicities, further necessitating minimizing the differences between observed and simulated relations \citep{hirschmann2023}.

In this work, we use galaxies from the RUBIES survey (\citealt{degraaff2025}) to construct a mass--metallicity relation at $z\sim$4. We measure metallicities using multiple strong-line diagnostics. We then generate artificial galaxy models and sample from these according to a star forming main sequence, the empirical RUBIES selection function, and measurability of metallicity according to the strength of emission lines. We posit a parameterized form of the MZR, and iteratively compare this to our RUBIES MZR, converging upon a best-fit relation. 

In Sec.~\ref{sec:data} we describe both the RUBIES and \texttt{prospector} samples. In Sec.~\ref{subsec:line_ratios} we characterize our line ratio measurement from RUBIES spectra and active galactic nuclei (AGN) removal, and in Sec.~\ref{subsec:masses} we describe the measurement of stellar masses. In Sec.~\ref{subsec:metallicity_mcmc} we detail the MCMC used to measure metallicities for each RUBIES object. In Sec.~\ref{sec:forward_modeling} we explain our MZR forward modeling process, including the \texttt{prospector} synthetic spectra generation, the star formation rate sampling, the empirical RUBIES selection function, metallicity measurability, and the likelihood function and priors. Finally, in Sec.~\ref{sec:results}, Sec.~\ref{sec:discussion}, and Sec.~\ref{sec:conclusion}, we present our results, discuss, and conclude, respectively. Throughout this paper we assume the following cosmology: H${_0}$ = 69.32 km s$^{-1}$ Mpc$^{-1}$, $\Omega_M$ = 0.2865, and $\Omega_\Lambda$ = 0.7135. To compute stellar masses, we assume a \citet{Chabrier2003} initial mass function.

\begin{figure*}
    \centering
    \includegraphics[width=0.9\linewidth]{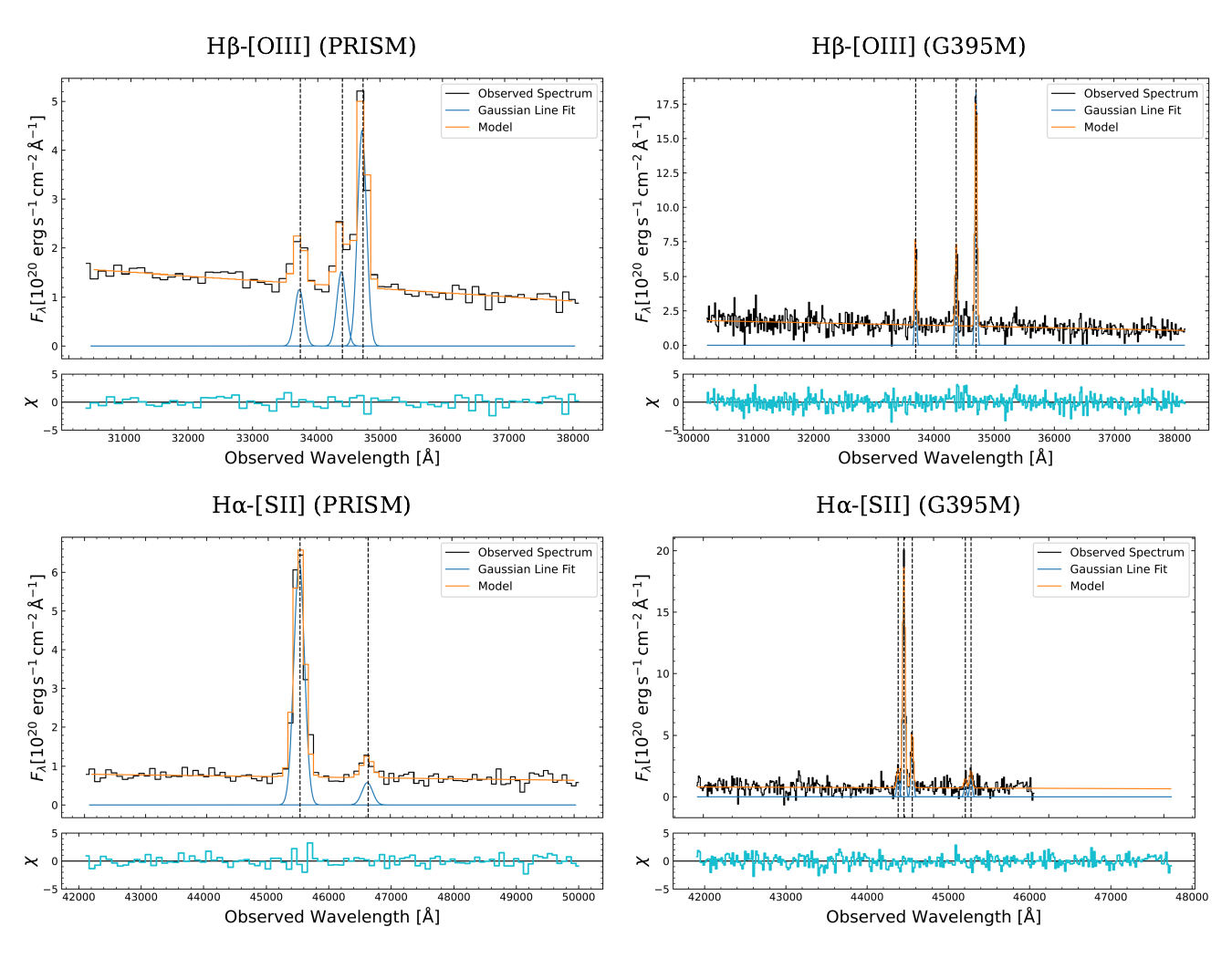}
    \caption{The emission line and continuum fits for RUBIES ID 37791. The raw spectrum is shown in black, the emission line fits in blue, and the fit reduced to the observed resolution in orange. The four panels correspond to the permutations of grating (PRISM and G395M) and complex (\hb-\oiii\ and \ha-\nii-\sii). The \nii\ lines are fixed to a 1:2.94 ratio. In the PRISM spectra, the \sii\ doublet is fit with a single Gaussian, and \ha\ is blended with \nii. We do not use this blended \ha\ measurement for any of our science cases. Subplots below each panel show residuals. Due to an absolute flux offset between measured PRISM and G395M fluxes, we adjust the measured G395M \ha\ flux when comparing with PRISM \sii\ fluxes; see Sec.~\ref{subsec:line_ratios}. We find that the fitting mechanism is able to successfully reproduce the observed data.}
    \label{fig:spectrum}
\end{figure*}

\section{Data}
\label{sec:data}

In this paper we utilize JWST Near InfraRed Spectrograph (NIRSpec) PRISM and G395M spectra from the Red Unknowns: Bright Infrared Extragalactic Spectroscopic (RUBIES; \citealt{degraaff2025}) Survey. We also generate \texttt{prospector} \citep{johnson2021} spectra across a wide range of parameter space to forward model the mass--metallicity relation. 

The RUBIES Survey (program ID 4233) is a JWST NIRSpec survey of NIRCam-selected galaxies. Targeting priority (outside of primary targets) was assigned according to three parameters: F444W flux, F150W-F444W color, and photometric redshift; see \citet{degraaff2025} for full details. Generally, weight was assigned according to inverse number density in this three-dimensional number space, meaning RUBIES prioritizes bright, red galaxies at high redshifts. This simple weighting scheme allows us to easily model the RUBIES selection function. The RUBIES sample consists of over 4000 spectroscopic targets, see \citet{degraaff2025} for details. In this work, we utilize the subsample of RUBIES that was observed with both the PRISM and G395M gratings on NIRSpec, consisting of around 3000 galaxies. The exposure time of 48 minutes for each galaxy enables the measurement of the most important Balmer and forbidden emission lines. The spectra were reduced using v4 of the \texttt{msaexp} pipeline \citep{brammer2023}, see also Brammer et al. in prep.

\section{Analysis}
\label{sec:rubies_mzr}

In this section we describe the measurement of emission lines from PRISM and grating spectra, the removal of AGN using those emission lines, the measurement of stellar masses, the determination of a metallicity from the emission lines, and the weighting of each object as a function of our ability to measure a metallicity in that regime of mass--metallicity parameter space. 

\subsection{Measuring Line Ratios}
\label{subsec:line_ratios}

Metallicities are most reliably calculated from the strength of auroral emission lines and electron temperatures, this being known as the ``direct method'' of metallicity inference (\citealt{izotov2006}, \citealt{curti2017}). These auroral lines, however, are often weak, especially in high-redshift spectra \citep{maiolino2019}. It is therefore more common to employ diagnostic relationships between strong emission lines and metallicities that have been tuned to direct method metallicities. In this work, we use the metallicity-sensitive R3 and S2 line ratios for reasons outlined in Sec.~\ref{subsec:metallicity_mcmc}. These line ratios are defined as:

\begin{equation}
    \text{R}3 = \text{log}\left(\frac{[\mathrm{OIII}]\lambda5008}{\mathrm H\beta}\right)
\end{equation}
\begin{equation}
    \text{S}2 = \text{log}(\frac{[SII]\lambda\lambda6717,6731}{H\alpha})
\end{equation}

These ratios necessitate the measurement of \hb, \oiiiboth, \ha, \nii, and \sii. We use the \texttt{emcee} package \citep{foreman-mackey2013} to fit single-component Gaussians to these lines using the methodology described in \citet{degraaff2025} after the redshifts have been fixed (over 94\% of the sample have measured redshifts; the remainder are not used). Most objects have both a low-resolution PRISM and medium-resolution G395M spectrum, and we fit each spectrum separately. To avoid needing to fit the complicated stellar continuum over the entire observed wavelength range, we fit lines one ``complex'' at a time: the \hb-\oiii\ complex and the \ha-\nii-\sii\ complex. We fit the \hb\ and \oiii\ (with the ratio fixed to 1:2.98) lines in both the PRISM and G395M spectra. We fit \ha\ \& \nii\ and the \sii\ doublet each as single Gaussians in the PRISM spectra, meaning that \ha\ is blended with \nii\ and thus not used for this science case. In the G395M spectra, \ha\ and \nii\ are fit separately (with the \nii\ doublet ratio fixed to 2.94) and \sii\ is fit with two Gaussians. 

In this process, Gaussians are generated with a given center, height, and width, then downsampled to the resolution of the observed spectrum and convolved with the LSF of an idealized point source before being compared to the observed spectrum (see Section 3.3 in \citealt{degraaff2025}). The continuum is fit as a first-degree polynomial simultaneously with the Gaussian emission lines. We do not account for underlying stellar absorption in e.g., \hb, since line fluxes are much brighter than expected absorption. An example fit is shown in Fig.~\ref{fig:spectrum} for RUBIES ID 37791. The observed spectrum is shown in black, the emission line fits in blue, and the line and continuum fit reduced to the resolution of the observed spectrum in orange. The figure consists of four panels: permutations of the two gratings used by RUBIES (PRISM and G395M), and the two line complexes necessary for this science case (\hb-\oiii\ and \ha-\nii-\sii). Subplots below each panel show residuals.

There exists an offset in absolute flux in NIRSpec spectra between fluxes measured from PRISM and from G395M spectra on the order of 20\% that appears to be constant and independent of wavelength \citep{degraaff2025}. We therefore derive a corrective factor (1.139, with scatter of 0.290 in flux units) by comparing the PRISM and G395M \oiiifive\ fluxes (see \citealt{degraaff2025}). We then, when using PRISM \sii\ fluxes for our metallicity calculation, create a ``pseudo'' PRISM \ha\ flux by applying this corrective factor to the G395M \ha\ measurement. We also inflate the pseudo-\ha\ uncertainty according to the scatter in this \oiiifive\ relation. This allows a valid S2 line ratio to be calculated for PRISM \sii\ fluxes.

\subsubsection{Removing AGN}
Ionization from active galactic nuclei (AGN) can contribute significantly to the strengths of emission lines in a spectrum, invalidating the calculation of a metallicity as line ratio values will be affected. It is thus necessary to remove suspected AGN from our sample. We use the \citet{kewley2001} AGN/SF separation curve and measurements of R3 and N2 to separate galaxies based on dominant ionization mechanism, though we caution that at low metallicities and high ionization parameters, AGN and galaxies have nearly complete overlap, and this overlap is also a function of metallicity \citep{cleri2025}. We use this curve on an ``opt-in'' basis: \nii\ is not successfully measured in all galaxies. Only objects with well-measured \nii\ that are determined to be driven by an AGN according to this diagnostic are removed; those for which \nii\ cannot be measured are not removed. 23 galaxies are flagged as AGN and are removed from our sample, primarily above $z\sim$5. The efficacy at higher redshifts of AGN diagnostics that have been calibrated in the local Universe remains uncertain; since the bulk of our sample lies between $3<z<5$, we do not expect this to affect our results. We also reject two ``Little Red Dots'' as characterized in \citet{hviding2025}, leaving 193 galaxies in our sample.

\subsection{Stellar Masses}
\label{subsec:masses}

Stellar masses are inferred following the methodology outlined in \citet{Wang2024:sps}, using JWST/NIRCAM + HST photometric data with redshifts fixed to the corresponding spectroscopic redshifts. We employ the \texttt{\texttt{prospector}} Bayesian inference framework \citep{johnson2021}, utilizing the MIST stellar isochrones \citep{Choi2016,Dotter2016} and the MILES stellar library \citep{Sanchez-Blazquez2006} as implemented in FSPS \citep{Conroy2010}. 

The stellar initial mass function is taken from \citet{Chabrier2003}. The continuity SFH is modeled using a non-parametric approach, defined by the mass formed in 7 logarithmically spaced time-bins (\texttt{prospector}-$\alpha$; \citealt{Leja2017}). Priors on the stellar mass and SFH, designed to optimize the photometric inference of deep JWST surveys, are adopted from \citet{Wang2023:pbeta}. Sampling is performed with the dynamic nested sampler \texttt{dynesty} \citep{Speagle2020}, with model generation accelerated by a neural net emulator \citep{Mathews2023}. We note that outshining could potentially bias stellar masses to lower values if light is dominated by younger stellar populations.

\subsection{Inferring Metallicity}
\label{subsec:metallicity_mcmc}

\begin{figure}
    \centering
    \includegraphics[width=\linewidth]{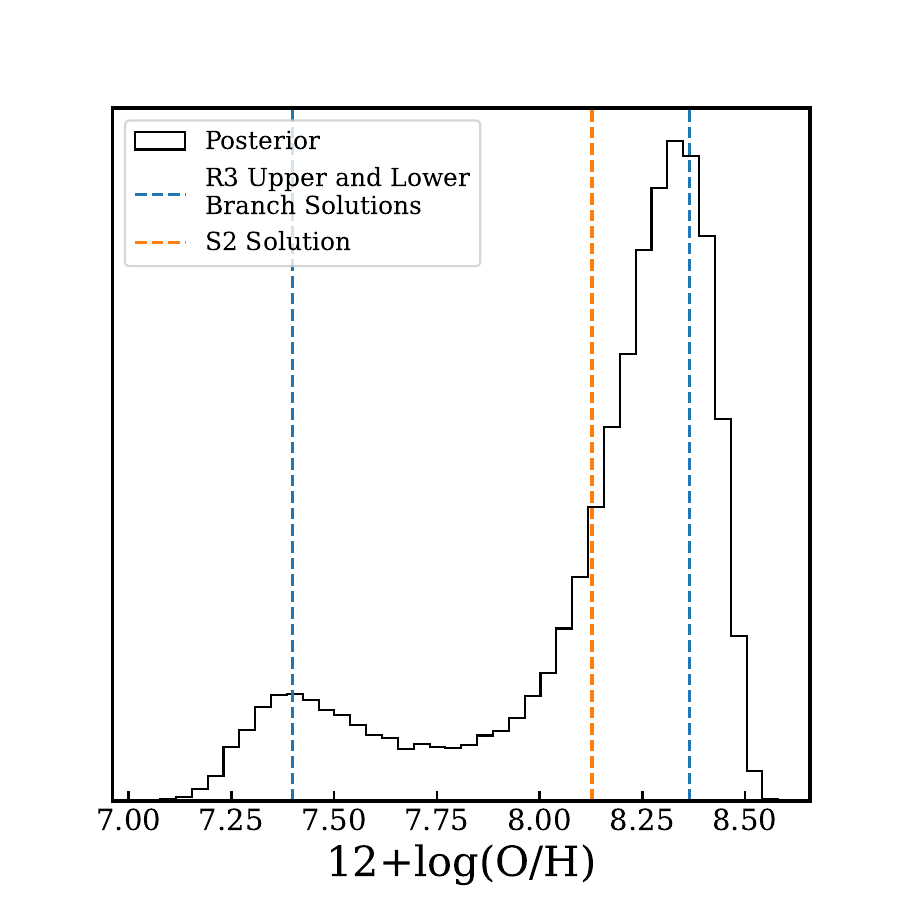}
    \caption{Metallicity MCMC posterior for RUBIES object 37791. The use of multiple diagnostic metallicity relations does not always completely break the metallicity degeneracy; the double-Gaussian nature of the converged posterior is evident in this figure. We sample from the entire metallicity posterior in our fitting procedure as opposed to assigning a single metallicity value to preserve this uncertainty. The double-branched R3 solutions and the S2 solution are shown in blue and orange, respectively.}
    \label{fig:posterior}
\end{figure}

As stated in Sec.~\ref{subsec:line_ratios}, we use the R3 and S2 line ratios to compute metallicity. We choose these ratios for two primary reasons. First, these line ratios exclusively use lines close together in wavelength space, avoiding complications with the NIRSpec and DJA absolute flux calibration \citep{maseda2023}, as well as differential dust reddening. Second, the aforementioned turnover for these relations exists at different metallicities, such that breaking the degeneracy is possible and the MCMC is able to converge to metallicities. We note, however, that all metallicity diagnostics may evolve with redshift; see, e.g., \citet{hirschmann2023b}. 

However, relations between line ratios and metallicity are known to be double-valued because of a lack of metal ions at low metallicities, as well as an inability for the metal ions to radiatively de-excite due to high ion densities at high metallicities \citep{maiolino2019}. We use multiple diagnostic relations to break this degeneracy. For objects with well-measured R3 and S2 (see Sec.~\ref{subsec:line_ratios}), we utilize a Markov Chain Monte Carlo (MCMC) to calculate metallicities via a Bayesian framework as described below. We require that the two line ratios have S/N$>$1. This cut aligns well with our visual inspection of spectra and their fits and is used for maximum completeness. 

Because of the availability of both PRISM and G395M spectra for the majority of our objects, we are afforded four permutations of inferred metallicities, corresponding to combinations of PRISM and G395M \sii\ and PRISM and G395M \oiiifive\ and \hb\ (only G395M \ha\ is used to avoid blending with \nii). We run the metallicity MCMC a maximum of four times for each object, corresponding to the availability of line flux measurements. For objects with multiple metallicity measurements, we prioritize in the following order: 
\begin{itemize}
    \setlength{\itemsep}{1pt}
    \item Objects with PRISM R3 and G395M S2 measurements,
    \item Objects with G395M R3 and G395M S2 measurements,
    \item Objects with PRISM R3 and PRISM S2 measurements,
    \item Objects with G395M R3 and PRISM S2 measurements.
\end{itemize}
We always prefer measurements where the \sii\ doublet is resolved; within that, we prefer measurements of PRISM R3 because the medium-resolution spectra have a noisier continua. We choose to use one of the metallicity posteriors as opposed to averaging multiple to retain the double-Gaussian shape of our posteriors.

The MCMC is modeled off that of \citet{wang2017} and \citet{wang2019} and uses the \texttt{emcee} package \citep{foreman-mackey2013} (see \citealt{lewis2024} for more details). It samples combinations of metallicity, nebular dust extinction, and dereddened \hb\ flux ($f_\mathrm{H\beta}^{\mathrm{dered}}$), which can be used to generate \oiiiboth, \ha, \hb, and \sii\ line flux ratios given a metallicity calibration and assuming case B recombination for the Balmer decrement. We describe this process in detail below. 

There are myriad sets of R3 and S2 metallicity calibrations available for use. We choose to use the \citet{nakajima2022} R3 and S2 relations and note that there exist other calibrations that may be used, e.g. \citet{sanders2024}. The calibrations of \citet{sanders2024} have the same shape as those of \citet{curti2017} and \citet{nakajima2022} save for an absolute offset in metallicity. This offset manifests as a downward shift in the normalization of our metallicities of $\sim$0.2 dex, which does not affect any of our qualitative conclusions. We then restrict our sample to objects with line ratios that fall in the calibrated range of the \citet{nakajima2022} diagnostics. 

We adopt a flat prior on metallicity (12~+~log(O/H) $\in$ [7.0, 8.9]) and on nebular attenuation (A$_V$ $\in$ [0, 6]), the latter using the nebular attenuation curve of \citet{cardelli1989}. No object converges to an extinction at the upper end of that scale; the range exists to allow a wide exploration of parameter space. We use the Jeffreys' prior distribution (\textit{prior} $\propto -\textrm{ln}(H\beta)$) for $f_\mathrm{H\beta}^{\mathrm{dered}}$ and limit its range to [0 to 10$^4$] in units of 10$^{-19}$ erg s$^{-1}$ cm$^{-2}$, corresponding to a star formation rate range of 0-2100 M$_{\odot}$  yr$^{-1}$ at our median redshift of $z=$3.8 using the conversion from \citet{kennicutt2012}. The likelihood function for our MCMC is given by
\begin{equation}
    L = -\sum\limits_{i} \frac{(f_i^{\mathrm{dered}} - R_i \cdot f_{\mathrm{H\beta}}^{\mathrm{dered}})^2}{\sigma_{f_i^{\mathrm{dered}}}^2 + \sigma_{R_i}^2 (f_{\mathrm{H\beta}}^{\mathrm{dered}})^2},
\label{eqn:chisq}
\end{equation}
where $f^{\mathrm{dered}}_i$ is the dereddened line flux of emission line \textit{i}, $\sigma_{f_i^{\mathrm{dered}}}$ is its measured uncertainty, $R_i$ is the ratio of the dereddened flux of emission line $i$ to the dereddened H$\beta$ flux, and $\sigma_{R_i}$ is the intrinsic scatter in the $R_i$--metallicity relation from \citet{nakajima2022} (where $\sigma_{R_i}=0$ for the Balmer lines because we assume no uncertainty in the case B recombination \ha/\hb\ line ratio).

The aforementioned double-branching of the R3 metallicity diagnostic often results in a double-peaked metallicity posterior; the S2 diagnostic helps to break the degeneracy in the sampling process, but not always completely. A novel aspect of this work is that we do not select a single metallicity value for each object. When fitting the MZR in subsequent sections, we sample from the full metallicity posteriors as opposed to assigning a single metallicity value to each galaxy. This ensures we do not neglect, e.g., the probability that a galaxy is a lower-branch metallicity object. We also include simultaneous independent draws from the stellar mass posterior, resulting in each object being represented as a ``cloud'' in mass--metallicity space, giving rise to a RUBIES mass--metallicity surface as we combine the mass and metallicity draws.

Fig.~\ref{fig:posterior} shows the metallicity posterior for RUBIES object 37791. The double-branched R3 metallicity solutions and S2 solution are shown in blue and orange, respectively. The use of multiple diagnostics results in a wider posterior, but more accurately captures the physical uncertainty of metallicity calculation. 

\begin{figure*}
    \centering
    \includegraphics[width=\linewidth]{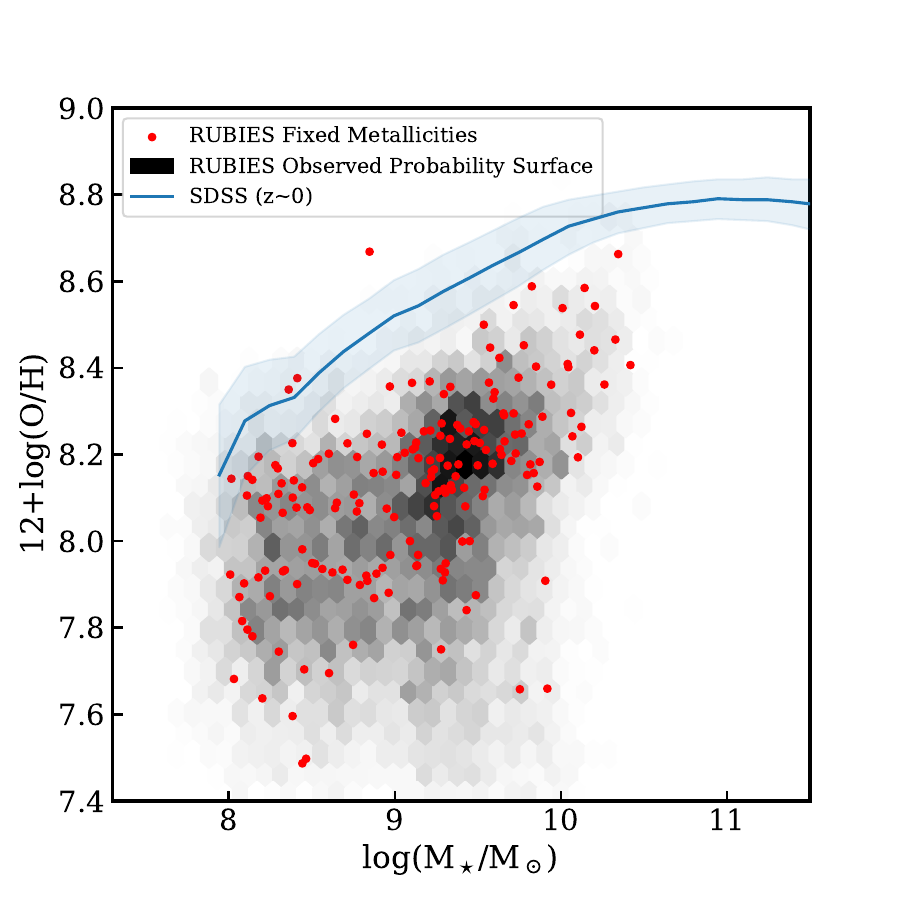}
    \caption{In red we show the RUBIES mass--metallicity points at $z=3-6$. We take the peak of the metallicity posterior distribution as the fixed metallicity for each galaxy. In grayscale we show the RUBIES MZR surface. Each galaxy is sampled 100 times from the metallicity posterior and Gaussian stellar mass uncertainties. The surface exists in two components: a flatter portion below log(M$_\star$/M$_\odot$)$\sim$9, and a positively-sloped portion above that, more akin to the expected behavior of low-redshift MZRs. The local Universe SDSS MZR from \citet{curti2020} is shown in blue.}
    \label{fig:mzr_surface}
\end{figure*}

Fig~\ref{fig:mzr_surface} shows in red the RUBIES mass--metallicity points, in which we assign a single metallicity (the peak of the metallicity posterior distribution) to each galaxy. The RUBIES mass--metallicity surface is shown in grayscale. Each RUBIES galaxy is sampled 100 times according to its metallicity posterior and Gaussian mass uncertainty; these comprise the RUBIES mass--metallicity relation surface. The surface appears to exist in two distinct portions: a lower-mass, flatter relation, and a higher-mass, positively sloped relation, closer to the correlation expected from low-redshift MZRs. The physical reasoning for this is explored below. Comparing the fixed metallicities with the metallicity surface highlights the necessity of considering full posteriors because of the bimodality inherent in the metallicity inference process.

\section{Forward Modeling the MZR}
\label{sec:forward_modeling}

In this section we describe the means by which we construct the $z\sim$4 MZR using RUBIES data. We begin by generating \texttt{prospector} models across a wide range of parameter space. We then sample these spectra according to a redshift-dependent star-forming main sequence, the RUBIES selection function, our ability to measure metallicity, and an iterative MZR using a fixed functional form. We then compare this MZR surface to our observations, using an MCMC to converge upon the best-fit parameters of the $z\sim$4 MZR and accounting for the physical and observational effects that can qualitatively change the shape of this relation. 

\begin{figure*}
    \centering
    \includegraphics[width=\linewidth]{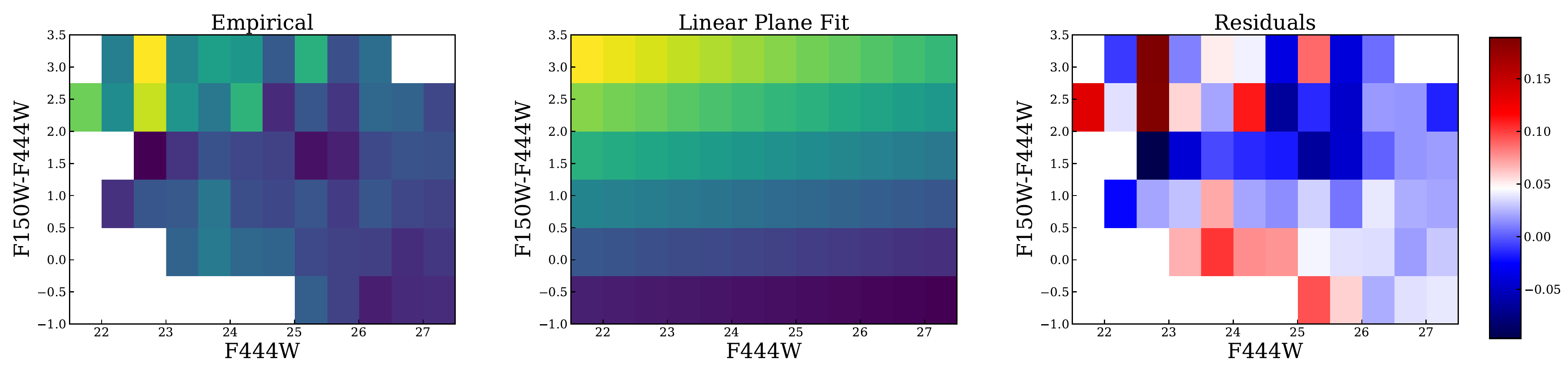}
    \caption{Our process of modeling the RUBIES selection function. The left panel shows the fraction of galaxies that have confirmed spectroscopic redshifts in bins of F444W and F150W-F444W space that were selected from a parent photometric catalog (\citealt{weibel2024}) for RUBIES targeting on a normalized scale. The middle panel shows our fit to this surface, linear in both magnitude and color. Both the left and middle panel are colored on a normalized scale (yellow being 1, dark blue being 0). The right panel shows the residuals from this fit, with no evidence of structure in the heavily-populated regions, suggesting that we have accurately parameterized the RUBIES selection function to first order. Fitting the selection function plane as opposed to using the raw fraction enables us to assign a weight to the regions of parameter space in which no RUBIES objects were targeted (e.g., the bottom left corner of the first panel).}
    \label{fig:selfunc}
\end{figure*}

\subsection{RUBIES Selection Function}
\label{subsec:selfunc}

In this section we empirically model the RUBIES selection function. We begin with the photometric parent catalog (\citealt{weibel2024}) from which RUBIES targets were selected, as well as the RUBIES catalog itself. We then bin these catalogs in F444W and F150W-F444W, as target selection weighting was set using these parameters. By plotting the fraction of objects in the target catalog that were selected for observation by RUBIES in each magnitude-color bin, we develop an empirical selection function surface. We verify that the F444W and F150W-F444W observation fractions are independent of redshift across the redshift range of our sample. 

We then fit this surface with a plane, linear in both F444W and F150W-F444W. This plane provides the selection function sampling probability for each \texttt{prospector} galaxy based on its F444W magnitude and color.

This process is illustrated in Fig.~\ref{fig:selfunc}. The left panel shows the empirically measured selection fraction as a function of 4-micron magnitude and color (on a normalized scale), the middle panel the linear planar fit to the fraction, and the right panel the residuals. We find that the linear fit to F444W and F150W-F444W accurately captures the selection function, as evidenced by the lack of structure in the right-most panel of the figure. We choose to fit the selection function plane as opposed to using the raw fractions because we expect the selection to be smooth in parameter space, but small number statistics prevents this from being realized in actuality.

\subsection{Prospector Model Generation}
\label{subsec:prospector_generation}

We generate a matrix of \texttt{prospector} (\citealt{johnson2021}) models with the goal of sampling observable parameter space to forward model the RUBIES mass--metallicity surface. \texttt{prospector} uses the Flexible Stellar Population Synthesis (FSPS; \citealt{Conroy2010}) to create a galaxy spectrum using given stellar population parameters by combining simple stellar populations with a star formation history. These parameters can either remain free, and defined by the user, or fixed. FSPS also includes self-consistent dust and nebular emission as well as intergalactic medium attenuation. For more details, see \citet{johnson2021}. 

When generating \texttt{prospector} data, we use the FastStepBasis simple stellar population (SSP), and we create both spectra and F150W and F444W photometry. We generate models by varying four physical parameters: redshift, stellar mass, metallicity, and star formation rate. The ranges of these parameters are shown in Tab.~\ref{tab:prospector}. We use a two-step star formation history, with the current star formation rate fixed over the past 50 Myr, and the earlier star formation rate fixed by the age of the galaxy according to its redshift and its stellar mass. Models are generated uniformly across these four parameters, with 50 steps for each parameter, resulting in 6,250,000 models. We do not generate models for the region of parameter space with objects of extremely low stellar mass and high star formation rate; the stellar mass accrued over the past 50 Myr would outweigh the input stellar mass. The bounds of these parameters are chosen to give a wide range of available parameter space across which to construct the MZR. The ionization parameter log(U) is set via a fixed relation according to the specific star formation rate by fitting the log(U)-sSFR relation in Fig. 11 of \citet{kaasinen2018}. The \texttt{prospector} photometry and spectra are reddened according to the star formation rate using the reddening-SFR relation found in Table 4 of \citet{garn2010}. Use of reddening-independent line ratios minimizes the importance of reddening emission line fluxes. The photometry, however, is more impacted: but we find in Sec.~\ref{subsec:impact} that the RUBIES selection function has little impact on the shape of the observed MZR; it is therefore not of significant importance the particular reddening relation we use. 

\begin{table}[htp]
    \centering
    \begin{tabular}{l|c|c}
        \hline 
       Parameter & Minimum & Maximum \\
       \hline
       log(Stellar Mass) & 8 & 11.5 \\
       log(SFR) & -0.5 & 3 \\
       Redshift & 3 & 7 \\
       12+log(O/H) & 6.5 & 9 \\
       \hline
    \end{tabular}
    \caption{Parameter ranges for our \texttt{prospector} generation. 50 objects are generated uniformly across each of the four parameters for a total of 6,250,000 models.}
    \label{tab:prospector}
\end{table}

\subsection{Star Formation Rate Sampling}
\label{subsec:sfrsampling}

The first step in the forward modeling process is developing a representative galaxy population by sampling according to star formation rate. We sample according to the redshift-dependent star-forming main sequence in Eqn. 28 of \citet{speagle2014}: sampling probabilities are assigned according to a Gaussian with a mean star formation rate as a function of stellar mass and redshift. The standard deviations of the star formation rates are set to 0.2 dex, in line with the findings of the ``true'' scatter. 

We do not sample according to stellar mass because our likelihood function compares metallicities binned by mass, so the relative number of objects in each mass bin is irrelevant for our forward modeling purposes. Additionally, because mass uncertainties are on the order of $\sim$0.1-0.2 dex, lower than the intrinsic scatter of the star formation main sequence, we are able to neglect this sampling. 

\subsection{Metallicity Measurability}
\label{subsec:measurability}

The ability of a galaxy's metallicity to be recovered is a function of physical parameters correlated with the metallicity itself. Galaxies with low masses and/or low star formation rates, that is, those with weaker emission lines, are less likely to have a measurable metallicity in our strong line approach. 

To account for this, we model metallicity measurability using our \texttt{prospector} models. We bin the objects in mass, metallicity, and redshift space, then measure the fraction of objects in each cell that pass the line ratio S/N test required of RUBIES objects themselves, described in Sec.~\ref{subsec:metallicity_mcmc}: line ratio S/N $>$ 1. \texttt{prospector} does not generate uncertainties on line fluxes, to approximate these, we set the measurement uncertainty on \texttt{prospector} emission lines to the minimum line flux that our emission line fitting code was able to accurately measure, a reflection of the RUBIES exposure time.

\begin{figure*}
    \centering
    \includegraphics[width=\linewidth]{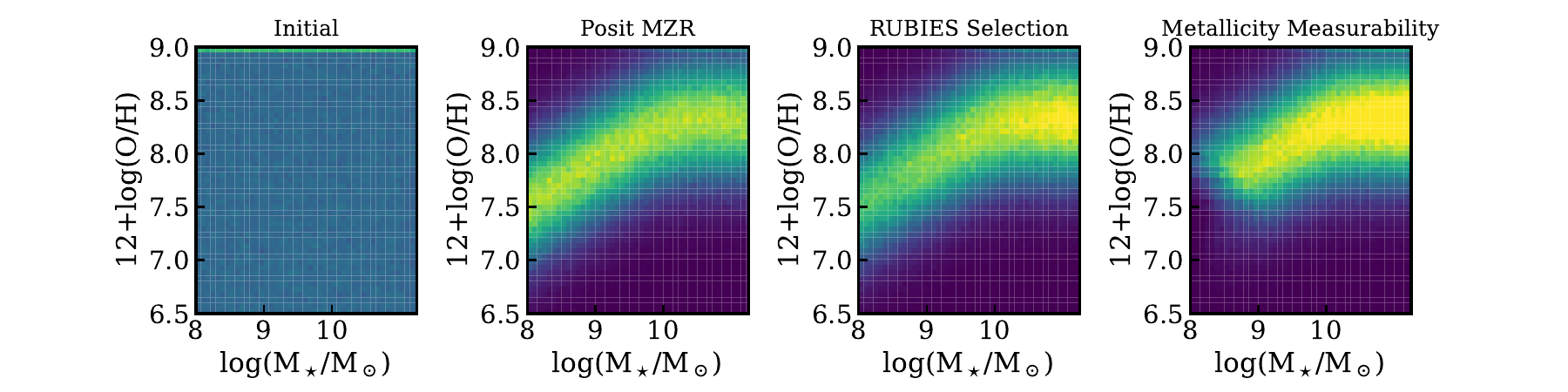}
    \caption{This figure illustrates the process by which the \texttt{prospector} matrix is sampled in order to compare to the RUBIES mass--metallicity surface. Each panel is shown on a normalized scale, such that yellow corresponds to 100\% of objects in that cell being selected, and dark blue 0\%. The first panel shows the initial \texttt{prospector} sample after star formation rate sampling. The second panel shows an example posited MZR in the form of Eqn.~\ref{eqn:mzr_fit} with mass-independent Gaussian scatter. The third panel shows the effect of the RUBIES selection function applied to the posited MZR, which is to prefer massive galaxies given RUBIES preferential targeting of bright objects. Finally, the fourth panel shows the effect of our ability to measure metallicity; namely emission line strength: dim galaxies with less luminous emission lines are undersampled. In our process, the selection function and metallicity measurability sampling happens prior to the MZR iteration; this figure shows the impact on the shape on a hypothetical given MZR.}
    \label{fig:prospector}
\end{figure*}

\subsection{Forward Modeling MCMC}
\label{subsec:forwardmodeling}

The star formation rate fractions, RUBIES selection function fractions, and metallicity measurability fractions comprise the {\it a priori} sampling performed on \texttt{prospector} galaxies. The next step is to posit a functional form of the MZR. We use the following form: 
\begin{equation}
    12 + \textrm{log(O/H)} = Z_0 -\frac{\gamma}{2} \textrm{log}\left(1 +\left(\frac{M}{M_0}\right)^{-2}\right),
    \label{eqn:mzr_fit}
\end{equation}
where $M_0$ is the characteristic turnover mass, $Z_0$ is the saturation metallicity that the MZR asymptotically approaches, $\gamma$ is the power law index of the MZR below $M_0$. We also include mass-independent Gaussian scatter $\sigma$ as a parameter. This equation is equivalent to the functional form used by \citet{curti2020} with $\beta$ fixed to 2. We fix $\beta$ because sample size restricts our ability to characterize the strength of the MZR turnover. Varying the fixed value of $\beta$ between 0 and 4 does not affect our outcomes. 

The process by which we sample from the \texttt{prospector} matrix in order to compare to the RUBIES mass--metallicity surface is shown in Fig.~\ref{fig:prospector}. The first panel shows our initial \texttt{prospector} sample in stellar mass and metallicity space after the star formation rate sampling. The second panel shows the \texttt{prospector} sample after sampling from the posited MZR with mass-independent Gaussian scatter. The third panel shows the impact of the RUBIES selection function. The effect is to weight the MZR more heavily towards high-mass galaxies, as RUBIES targeted objects bright at 4 microns. The rightmost panel shows the impact of our ability to measure metallicity; low-mass objects with weaker emission lines are removed in this case. In our process, the iterative MZR sampling happens after the selection function and measurability sampling. This figure is used to demonstrate the impact of those samplings on a given MZR. 

We run an MCMC for the $z_0$, $\gamma$, M$_0$, and $\sigma$ parameters. Our likelihood function operates as follows: An MZR is created with these values according to Eqn.~\ref{eqn:mzr_fit}, and the \texttt{prospector} matrix is sampled according to the parameters. We then slice this \texttt{prospector} MZR into stellar mass bins of width 0.15 dex, and compute means and standard deviations in each bin. While the imposed mass--metallicity relation is Gaussian, we verify that the selection function and metallicity measurability sampling do not change the shape of the distribution in each mass bin; that is, the metallicity distribution in each mass bin post-sampling is still Gaussian. Next, for each of the RUBIES galaxies, each of the 100 draws is assigned to its respective mass bin. The likelihood for each RUBIES galaxy is assigned according to the equation 
\begin{equation}
    L = -\Sigma_i \langle \left( \frac{Z_{p,j} - Z_{R_{i,j}}}{\sigma_{P_j}} \right)^2 + \textrm{log}(2\pi \sigma^2_{P_j}) \rangle_j,
    \label{eqn:likelihood}
\end{equation}
where $Z_{R_{i,j}}$ is the metallicity for the $j$th draw of the $i$th RUBIES galaxy, and $Z_{p_j}$ and $\sigma_{P_j}$ are the mean and standard deviations of the \texttt{prospector} metallicities in the mass bin assigned to that $j$th draw, respectively. That is, the Gaussian-weighted distance of each draw of each RUBIES object from the mean \texttt{prospector} metallicity in that mass bin is computed. These values are then averaged across the 100 draws to give a likelihood for each RUBIES galaxy, which are then summed in log space to give our final likelihood. We use flat priors for this MCMC, with $z_0$ allowed to vary between 8.4 and 9, $\gamma$ between 0 and 2, M$_0$ between 11 and 13, and $\sigma$ between 0.1 and 1. 

\section{Results}
\label{sec:results}


Our MCMC converged well for the $\gamma$ and $\sigma$ parameters, but only converged widely for the $z_0$ and M$_0$ parameters, with the latter being above the mass range of the RUBIES sample. This implies that RUBIES captures the positively-correlated region of the MZR well, but does not reach high enough stellar masses to accurately constrain the metallicity turnover and asymptote, but see Sec.~\ref{subsec:future_work} for a brief discussion. In the following sections, we use the well-converged values of the $\gamma$ and $\sigma$ parameters, and the highest-probability values of the M$_0$ and $z_0$ parameters. These values are given in Table \ref{tab:parameters}.

\begin{table}[htp]
    \centering
    \begin{tabular}{l|c}
        \hline 
       Parameter & Value \\
       \hline
       $z_0$ & 8.45$_{-0.05}^{+0.09}$ \\
       $\gamma$ & 0.13$_{-0.02}^{+0.02}$ \\
       M$_0$ & 12.80$_{-0.53}^{+0.40}$ \\
       $\sigma$ & 0.29$_{-0.01}^{+0.01}$ \\
       \hline
    \end{tabular}
    \caption{Converged MZR parameters according to Eqn.~\ref{eqn:mzr_fit} for our forward modeling MCMC. The $\gamma$ and $\sigma$ parameters are well-converged, whereas the M$_0$ and $z_0$ parameters have highest-probability values but are not well converged and thus are not shown with uncertainties. This implies the RUBIES MZR well captures the lower-mass, positively correlated portion of the MZR, but does not reach high enough stellar masses to constrain the turnover and asymptotic region.}
    \label{tab:parameters}
\end{table}

\begin{figure*}
    \centering
    \includegraphics[width=\linewidth]{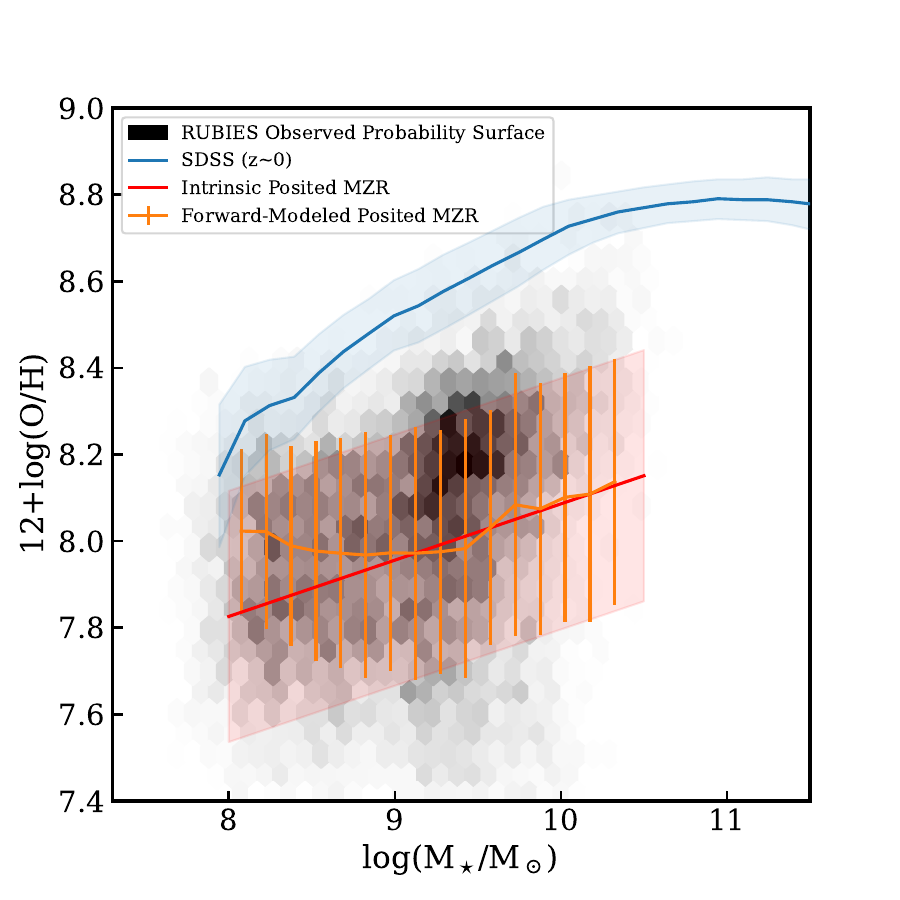}
    \caption{The RUBIES MZR at z=3-6. The RUBIES mass--metallicity surface is shown in grayscale hexbinning alongside the \citet{curti2020} SDSS $z\sim$0 MZR in blue. The best-fit MZR that the forward modeling MCMC converged to is shown in red. In orange we show this MZR applied to our \texttt{prospector} sample, including the effects of the RUBIES selection function and metallicity measurability. We find that these effects significantly impact the observed slope of the MZR below log(M$_\star$/M$_\odot$)$\sim$9.5, with the observed MZR appearing far flatter than the intrinsic one. Having measured the MZR without accounting for observational effects would lead to incorrect conclusions about the shape of the relation at these redshifts, demonstrating the necessity of our forward-modeling framework.}
    \label{fig:mzr_fit}
\end{figure*}

The RUBIES MZR is shown in Fig.~\ref{fig:mzr_fit}. As in Fig.~\ref{fig:mzr_surface}, we show the RUBIES mass--metallicity surface as grayscale hexbins and the SDSS $z\sim$0 relation from \citet{curti2020} as a blue line. The converged input MZR and its scatter is shown in red. That same mass--metallicity relation is shown applied to the \texttt{prospector} sample and after selection function and metallicity measurability in orange.  

Our primary finding is that the uncorrected MZR appears far flatter when the selection function and metallicity measurability biases are not accounted for (orange line); we explore the individual impacts of these effects in the following section. We find the slope of the MZR below log(M$_\star$/M$_\odot$)$\sim$9.5 is strongly affected by the RUBIES selection, flattening the relation as a whole. The normalization of the MZR is largely unchanged by these effects. 

The corrected RUBIES MZR, even when these biases are taken into account, is flatter than the local Universe MZR. The relation lies approximately 0.7 dex below the $z\sim$0 relation at log(M$_\star$/M$_\odot$)$\sim$10.5, and 0.4 dex below the $z\sim$0 relation at log(M$_\star$/M$_\odot$)$\sim$8. 

\section{Discussion}
\label{sec:discussion}
In this section we explore the impact of each step of the forward modeling process on the measured MZR before comparing our results to observational literature MZRs and discussing future work. 

\subsection{The Impact of Each Forward Modeling Step}
\label{subsec:impact}

\begin{figure}
    \centering
    \includegraphics[width=\linewidth]{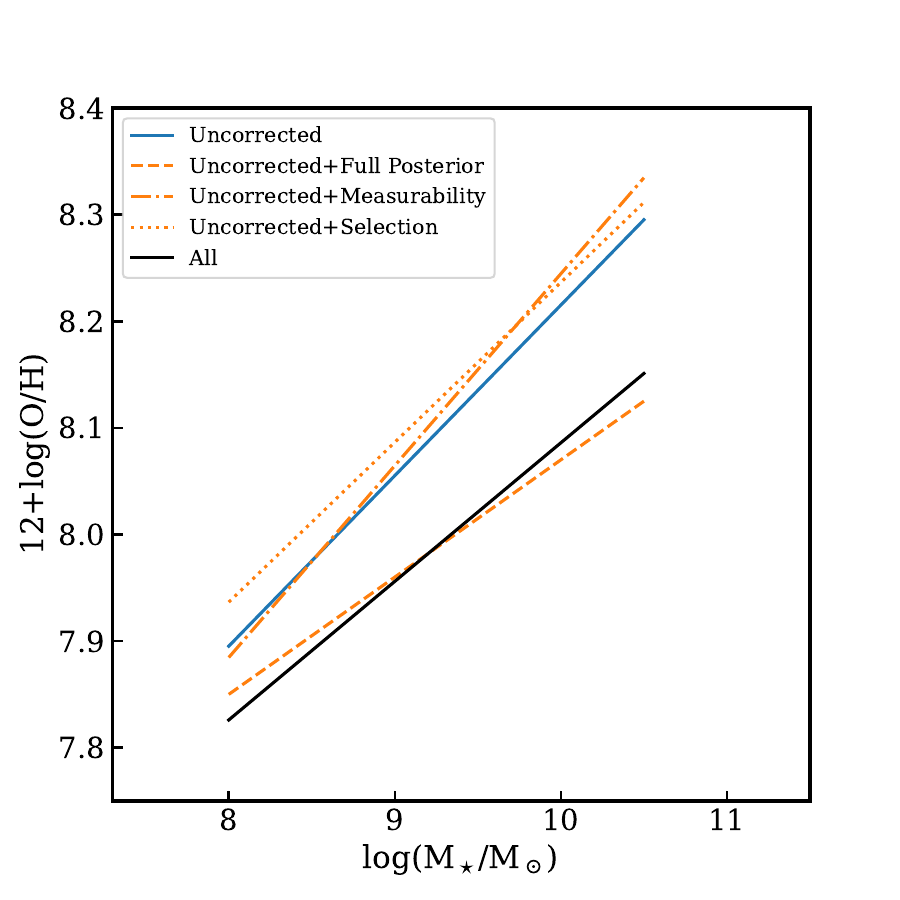}
    \caption{The impact of the various observational effects upon the MZR we accounted for in this work. All MZRs are fit using the likelihood function described in Sec.~\ref{subsec:forwardmodeling}. The ``uncorrected'' fit is shown as a blue dashed line. The ``uncorrected + resampled'' fit, which includes the effect of sampling from the metallicity posterior as opposed to assuming a single, fixed value, is shown as an orange dashed line. This fit has a lower slope and normalization than the uncorrected fit. The ``uncorrected + measurability'' fit, which includes the impact of metallicity measurability described in Sec.~\ref{subsec:measurability}, is shown as an orange dot-dashed line. This fit has a slightly higher slope than the uncorrected fit as low-mass, low-metallicity galaxies with weak emission lines are included in the original sample. The ``uncorrected + selection'' fit, which includes the impact of the RUBIES selection function as described in Sec.~\ref{subsec:selfunc}, is shown as an orange dotted line. This fit has a higher normalization than the uncorrected fit as dimmer galaxies with lower star formation rates and higher metallicities at fixed stellar mass are included. The intrinsic scatter on the MZRs, which is generally unchanged by the applied correction effects, is omitted for clarity. This MZR has a similar slope to the RUBIES unadjusted ``uncorrected'' MZR, but is $\sim$0.2 dex lower in normalization.}
    \label{fig:mzr_effect}
\end{figure}

Here we explore the impact on the MZR of three different effects we have accounted for in this work: metallicity diagnostic double branching, metallicity measurability, and the RUBIES selection function. We begin by fitting a ``basic'' MZR using the same likelihood equation (Eqn.~\ref{eqn:likelihood}), but without any of these effects accounted for. We then reintroduce these steps, one at a time, to this uncorrected fit to explore the effect on the shape of the MZR. 

As discussed in Sec.~\ref{subsec:metallicity_mcmc}, metallicity-strong line ratio diagnostic relations are often double-branched in that a given observed line ratio can be attributed to two values of a metallicity. While our metallicity MCMC begins to break this degeneracy by utilizing two metallicity diagnostics, we sample from the metallicity posterior instead of assigning single metallicity values to preserve the uncertainty inherent in this process. 

To explore the impact of this choice on our MZR, we fit two Gaussians to the posteriors using the \texttt{lmfit} package \citep{newville2014}, with varying centers, widths, and amplitudes. We then take the Gaussian associated with the peak of the posterior distribution as the fiducial metallicity, as opposed to that with the largest integrated area, as this method best recreates the metallicity-line ratio diagnostic relations described above. The 16th and 84th percentiles of the Gaussian are taken as the metallicity uncertainty. 

The fit to this ``uncorrected + resampled'' MZR is flatter than the uncorrected MZR, with a lower normalization (0.1 dex at log(M$_\star$/M$_\odot$)$\sim$10.5). This is because the majority of RUBIES galaxies have a majority of their metallicity distribution on the upper branch, such that including the possibility of a lower-branch solution lowers the overall normalization of the relation. 

We also account for the ability of our MCMC to measure metallicities, which is largely dependent on the strength of a galaxy's emission lines. This is described in detail in Sec.~\ref{subsec:measurability}. The primary impact of this is to account for the low-mass, low-metallicity galaxies that were removed from the observed sample because of their weak emission lines. This moderately increases the slope of the MZR fit, but leaves the normalization unchanged. 

The RUBIES selection function is accounted for as described in Sec.~\ref{subsec:selfunc}. Since RUBIES targets bright, red galaxies (see \citealt{degraaff2025} for full details), the impact of accounting for the selection function is to remove from our synthetic sample the low-mass galaxies with weaker emission lines. This marginally increases the overall normalization of the MZR, but leaves the slope unchanged. Even the impact on the normalization of this effect is minor: the F444W magnitude component of the RUBIES selection function translates at these redshifts to an effective stellar mass selection; the mass-independence of our forward modeling likelihood function renders this moot. The F150W-F444W color selection here translates to an effective rest-frame B-I color selection, which has only a weak dependence on metallicity, so the RUBIES selection function does not significantly impact the shape of our observed mass--metallicity relation. 

All of these effects combine to create an MZR that has a lower normalization and slightly lower slope than the ``uncorrected'' MZR. This means that not accounting for these effects would result in measuring a mass--metallicity relation that has a higher normalization than is true. 

These effects are captured in Fig.~\ref{fig:mzr_effect}, which displays the MZR fits to each of these perturbations to the ``uncorrected'' MZR fit. The uncorrected fit is shown as a blue dashed line, the ``uncorrected + resampling'' fit as an orange dashed line, the ``uncorrected + measurability'' fit as an orange dot-dashed line, the ``uncorrected + selection'' fit as an orange dotted line, and the combination of these effects as a black solid line. This relation has a similar slope to the RUBIES ``uncorrected'' MZR, but is $\sim$0.2 dex lower in normalization.

\subsection{Comparison to Literature MZRs}
\label{subsec:comparison}

\begin{figure*}
    \centering
    \includegraphics[width=\linewidth]{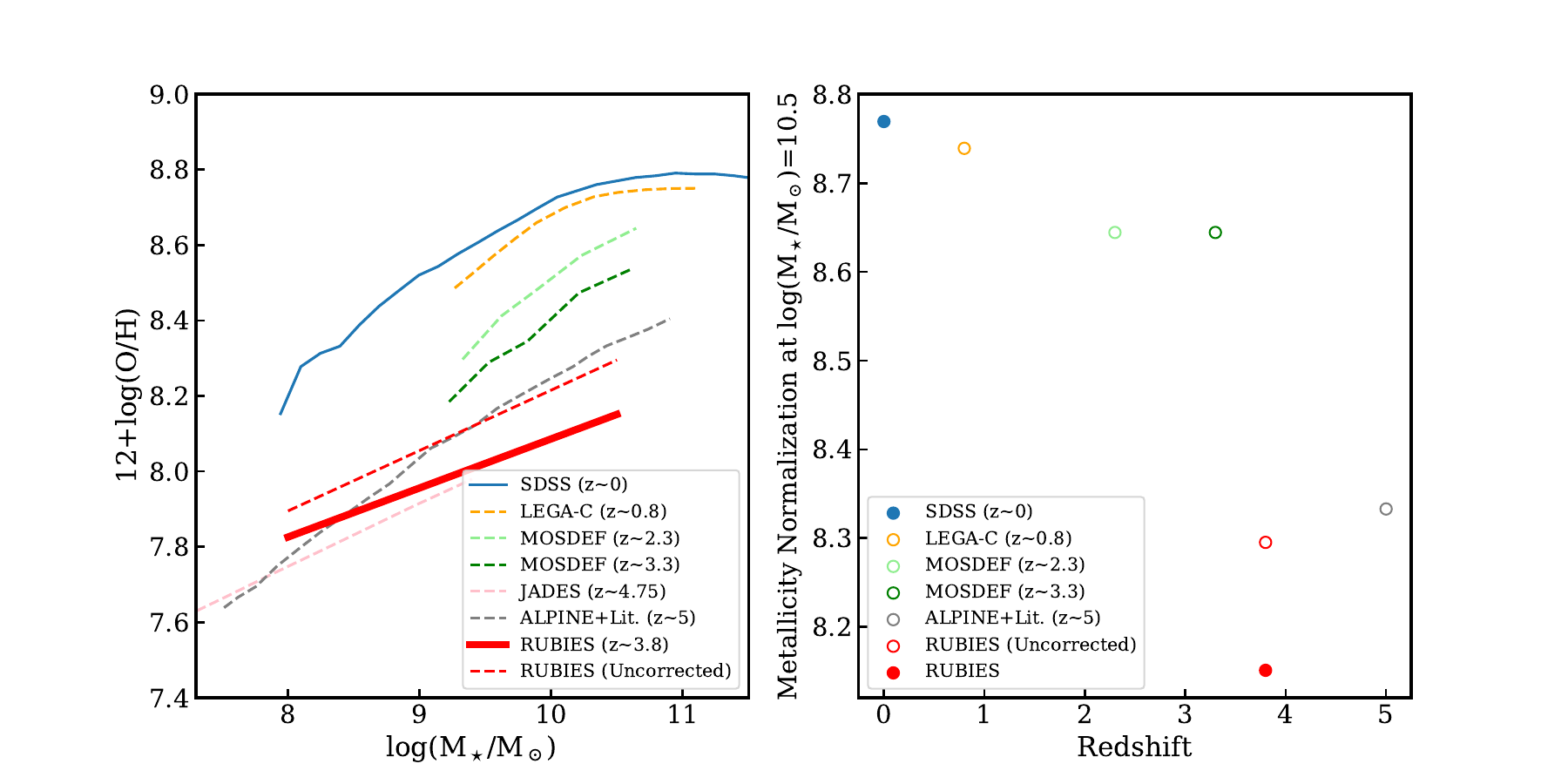}
    \caption{Left panel: A comparison of the RUBIES corrected and uncorrected MZRs to several literature observational MZRs. The local Universe SDSS relation is shown in blue, the intermediate redshift LEGA-C and MOSDEF relations in orange, light green, and dark green respectively, and the higher redshift JADES, ALPINE, and RUBIES relations in pink, grey, and red, respectively. Only the SDSS and corrected RUBIES relations are shown as solid lines to highlight their representativeness, the former because the effective stellar mass limit of SDSS is well below that considered in this work, the latter because of the attempts made in this paper to correct for biases described above. Right panel: The evolution of the normalization of the MZR at log(M$_\star$/M$_\odot$)=10.9, using the same colors as in the left panel. Also as in the right panel, only SDSS and the corrected RUBIES points are shown as filled circles. We find that the RUBIES points, particularly the corrected value, lie below the trend formed by the other surveys. We attribute this to our metallicity posterior sampling, and leave drawing conclusions about galaxy chemical evolution to a future work that has adjusted metallicity surveys at intermediate redshifts.}
    \label{fig:comparison}
\end{figure*}

We here compare both the corrected and uncorrected RUBIES mass--metallicity relation to several observational literature MZRs across redshift. We compare to the $z\sim$0 MZR of SDSS \citep{curti2020}, the $z\sim$0.8 MZR of LEGA-C \citep{lewis2024}, the $z\sim$2.3 and $z\sim$3.3 MZRs of MOSDEF \citep{sanders2021}, the $z\sim$4.75 MZR of JADES \citet{curti2024}, and the $z\sim$5 MZR of ALPINE and literature metallicities \citep{faisst2025}. 

The left panel of Fig.~\ref{fig:comparison} shows these relations alongside both RUBIES MZRs. Only the SDSS and corrected RUBIES MZRs are shown as solid lines to emphasize their representativeness; the former is representative because the effective luminosity or stellar mass limit at $z\sim$0 is well below that considered in this work, the latter because of the work described in this paper. 

The right panel of Fig.~\ref{fig:comparison} shows the evolution of the normalization of the MZR at log(M$_\star$/M$_\odot$)=10.9 as a function of redshift for the same set of surveys. As in the left panel, only the SDSS and corrected RUBIES values are shown as filled circles. The corrected RUBIES points lies below the trend formed by the other normalizations; we attribute this to our metallicity posterior sampling. We do not attempt here to characterize in depth the evolution of the mass--metallicity across cosmic time; we leave this to a future work which is able to consider the impacts of the selection functions and metallicity measurability of the aforementioned surveys. 

\subsection{Future Work}
\label{subsec:future_work}

The observational effects we accounted for in this work do not comprise the entirety of effects that could impact the shape of the measured MZR. This work is a first step in beginning to untangle the separation between the intrinsic and the observed mass--metallicity relation and the physical reasons behind that separation. Future work could include understanding biases in stellar mass inference, a more detailed treatment of survey selection effects, and a discussion of the fundamental metallicity relation as it relates to our correction for those selection effects \citep{laseter2025}. Future work could also include the metallicity inference as part of the forward modeling framework, developing a hierarchical formulation for the work done here, as well as considering the impact of using different metallicity line ratio diagnostics. 

More importantly, however, future work must attempt to address the biases discussed here in other surveys, with the goal of presenting a coherent and robust observed evolution of the mass--metallicity relation. Working to correct metallicity surveys at various redshifts will allow us to characterize the chemical evolution of galaxies in a more physically meaningful way than ever before---allowing accurate comparisons to hydrodynamical cosmological simulations and an understanding of the early epochs of the mass metallicity relation, as well as placing tight constraints on star formation and feedback processes.

An important extension to this work and others considering the evolution of the MZR, especially those using strong-line metallicity calibrations, is the existence of quiescent galaxies. As a galaxy quenches, and its emission lines become weaker, it evolves ``off'' of the measurable MZR. The relative dearth of quiescent galaxies at higher redshifts could partially explain our inability, then, to constrain the turnover and asymptote of the MZR. As the universe ages, the quiescent fraction at fixed stellar mass increases \citep{muzzin2013}. This leads to a decrease in the number of massive galaxies present on the MZR. This decrease, however, is combatted by a continuous increase in stellar mass of the galaxy population as a whole. This results in a complex interplay on the MZR in which the galaxy population moves upwards and rightwards, increasing in metallicity and stellar mass respectively, at a decreasing rate before becoming chemically saturated and ultimately disappearing from the diagram as star formation is quenched. Further work in simulations is necessary to trace individual galaxies across cosmic time to investigate the rate at which galaxies move across the diagram; JWST now allows these simulations to be compared to observations at the highest redshifts. 

\section{Conclusion}
\label{sec:conclusion}

In this work we characterized the RUBIES MZR at $z\sim$3-6. In doing so, we accounted for several observational affects that impact the shape of the measured MZR, including our survey selection function, our ability to measure metallicity based on emission line strengths, and the double-branching of line ratio metallicity diagnostics. 

We found that accounting for these effects had a qualitative impact on the shape of the RUBIES MZR. Our MZR had a lower normalization and lower slope than we would have otherwise measured had we ignored these effects. This demonstrates the importance of attempting to correct for observational biases, especially in metallicity studies.

Attempting to account for factors that introduce bias, or at least quantifying their effect on results, can affect the interpretation of the MZR. The consideration of environment and star formation rate as secondary parameters (the latter effectively extending the fundamental metallicity relation to high redshifts; see \citealt{sanders2021}) will also be necessary. Finally, pinning down the behavior of the MZR slope at early cosmic times will require further studies beyond cosmic noon, with wide, representative, and deep surveys. 

\section{Acknowledgments}

This work is based on observations made with the NASA/ESA/CSA James Webb Space Telescope. This material is based upon work supported by the National Science Foundation Graduate Research Fellowship under grant
No. 2137424 as well as work supported by NASA under Award No. 2025\_3-0, issued through the Wisconsin Space Grant Consortium, and JWST-GO-4233. Any opinions, findings, and conclusions or recommendations expressed in this material are those of the author(s) and do not necessarily reflect the views of the National Aeronautics and Space Administration. Support for program \#4233 was provided by NASA through a grant from the Space Telescope Science Institute, which is operated by the Association of Universities for Research in Astronomy, Inc., under NASA contract NAS 5-03127. MVM is supported by the National Science Foundation via grant AAG 2205519.  AdG acknowledges support from a Clay Fellowship awarded by the Smithsonian Astrophysical Observatory. TBM was supported by a CIERA Fellowship.
Part of the computations for this research were performed on the Pennsylvania State University's Institute for Computational and Data Sciences' Roar supercomputer.

\bibliographystyle{aa}
\bibliography{rubies_mzr.bib}
\end{document}

%% file: commands.tex
\newcommand{\fluxunit}{10$^{-20}$ erg s$^{-1}$ cm$^{-2} \rm \AA^{-1}$}
\newcommand{\Lya}{\hbox{{\rm Ly}$\alpha$}}
\newcommand{\lya}{\hbox{{\rm Ly}$\alpha$}}
\newcommand{\Ha}{\hbox{{\rm H}$\alpha$}}
\newcommand{\ha}{\hbox{{\rm H}$\alpha$}}
\newcommand{\Hb}{\hbox{{\rm H}$\beta$}}
\newcommand{\hb}{\hbox{{\rm H}$\beta$}}
\newcommand{\Hg}{\hbox{{\rm H}$\gamma$}}
\newcommand{\hg}{\hbox{{\rm H}$\gamma$}}
\newcommand{\Paa}{\hbox{{\rm Pa}$\alpha$}}
\newcommand{\paa}{\hbox{{\rm Pa}$\alpha$}}
\newcommand{\Pab}{\hbox{{\rm Pa}$\beta$}}
\newcommand{\pab}{\hbox{{\rm Pa}$\beta$}}

\newcommand{\heii}{\hbox{\ion{He}{2}}}
\newcommand{\cii}{\hbox{[\ion{C}{2}]}}
\newcommand{\ciii}{\hbox{[\ion{C}{3}]}}
\newcommand{\civ}{\hbox{\ion{C}{4}}}
\newcommand{\nii}{\hbox{[\ion{N}{2}]$\lambda 6548,6584$}}
\newcommand{\niv}{\hbox{[\ion{N}{4}]}}
\newcommand{\nv}{\hbox{[\ion{N}{5}]}}
\newcommand{\oi}{\hbox{\ion{O}{1}}}
\newcommand{\oii}{\hbox{[\ion{O}{2}]}}
\newcommand{\xoii}{\hbox{[\ion{O}{2}]$\lambda\lambda 3727,3729$}}
\newcommand{\oiii}{\hbox{[\ion{O}{3}]}}
\newcommand{\xoiii}{\hbox{[\ion{O}{3}]$\lambda\lambda 4960,5008$}}
\newcommand{\xoiiia}{\hbox{[\ion{O}{3}]$\lambda 5008$}}
\newcommand{\xoiiib}{\hbox{[\ion{O}{3}]$\lambda 4960$}}
\newcommand{\oiv}{\hbox{[\ion{O}{4}]}}
\newcommand{\neii}{\hbox{[\ion{Ne}{2}]}}
\newcommand{\neiii}{\hbox{[\ion{Ne}{3}]}}
\newcommand{\xneiii}{\hbox{[\ion{Ne}{3}] $\lambda 3870$}}
\newcommand{\nev}{\hbox{[\ion{Ne}{5}]}}
\newcommand{\xnev}{\hbox{[\ion{Ne}{5}] $\lambda 3347,3427$}}
\newcommand{\xneva}{\hbox{[\ion{Ne}{5}] $\lambda 3427$}}
\newcommand{\sii}{\hbox{[\ion{S}{2}]$\lambda 6717,6731$}}
\newcommand{\ariii}{\hbox{[\ion{Ar}{3}]}}
\newcommand{\ariv}{\hbox{[\ion{Ar}{4}]}}
\newcommand{\mgii}{\hbox{\ion{Mg}{2}}}
\newcommand{\feii}{\hbox{[\ion{Fe}{2}}]}
\newcommand{\fex}{\hbox{[\ion{Fe}{10}}]}

\newcommand{\vlya}{\hbox{$\Delta$v$_{Ly\alpha}$}}
\newcommand{\vciv}{\hbox{$\Delta$v$_{CIV}$}}
\newcommand{\wciii}{\hbox{W$_{\text{\sc C\,iii}]}$}~}
\newcommand{\wciv}{\hbox{W$_{\text{\sc C\,iv}}$}~}
\newcommand{\wlya}{\hbox{W$_{\text{Ly$\alpha$}}$}~}
\newcommand{\fesc}{\hbox{$f_{esc}$}}

\newcommand{\hst}{\textit{HST}}
\newcommand{\hstlong}{\textit{Hubble Space Telescope}}
\newcommand{\jwst}{\textit{JWST}}
\newcommand{\jwstlong}{\textit{James Webb Space Telescope}}
\newcommand{\romanlong}{\textit{Nancy Grace Roman Space Telescope}}
\newcommand{\HST}{{\it HST}}
\newcommand{\JWST}{{\it JWST}}
\newcommand{\spitzer}{{\it Spitzer}}
\newcommand{\Spitzer}{{\it Spitzer}}

\newcommand{\gndone}{z7\_GND\_42912}
\newcommand{\gndtwo}{z7\_GND\_16863}
\newcommand{\gndthree}{z7\_GND\_22483}

\newcommand{\cloudy}{{\sc Cloudy}}
\newcommand{\bpass}{{\sc bpass}}
\newcommand{\grizli}{\texttt{grizli}}
\newcommand{\pyneb}{\texttt{PyNeb}}

\newcommand{\snr}{\hbox{S/N}}

\newcommand{\logoh}{\hbox{$\log$(O/H)}}
\newcommand{\sphinx}{\hbox{SPHINX$^{20}$}}
\newcommand{\msun}{\hbox{M$_\odot$}}
\newcommand{\msol}{\hbox{M$_\odot$}}

\newcommand{\Hplus}{\hbox{H$^{+}$}}

\newcommand{\Oplus}{\hbox{O$^{+}$}}
\newcommand{\Otwoplus}{\hbox{O$^{2+}$}}
\newcommand{\Othreeplus}{\hbox{O$^{3+}$}}

\newcommand{\Netwoplus}{\hbox{Ne$^{2+}$}}

\newcommand{\njc}[1]{\textcolor{red}{\textbf{\texttt{#1}}}}